\documentclass[12pt]{article}
\usepackage{graphicx}
\usepackage[hang,small,bf]{caption}
\newcommand{\be}[3]{\begin{equation}  \label{#1#2#3}}

\newcommand{\ee}{ \end{equation}}
\newcommand{\ba}{\begin{array}}
\newcommand{\ea}{\end{array}}

\let\LARGE=\Large
\let\Large=\large

\begin{document}


\thispagestyle{empty}
\rightline{UPR-875-T}
\rightline{CALT-68-2262}
\rightline{CITUSC/00-009}
\rightline{hep-th/0002057}

\vspace{1truecm}

\centerline{\bf \LARGE 
Infra-red fixed points at the boundary
}
\bigskip

\vspace{1.2truecm}
\centerline{\bf Klaus Behrndt$^a$\footnote{e-mail: 
 behrndt@theory.caltech.edu}\quad
{\rm and} \quad
Mirjam Cveti{\v c}$^b$\footnote{e-mail: cvetic@cvetic.hep.upenn.edu}}
\vspace{.5truecm}
\centerline{$^a$ \em California Institute of Technology}
\centerline{\em Pasadena, CA 91125}

\bigskip

\centerline{\it CIT-USC Center For Theoretical Physics}
\centerline{\it University of Southern California}
\centerline{\it Los Angeles, CA 90089-2536}

\vspace{.3truecm}

\centerline{$^b$ \em Department of Physics and Astronomy}
\centerline{\em University of Pennsylvania, Philadelphia, PA 19104-6396}

\vspace{1truecm}


\begin{abstract}
Gauged supergravities (in four and five dimensions) with eight
supercharges and with vector supermultiplets have a unique
ultra-violet (UV) fixed point on a given physical domain ${\cal M}$ of
the space of the scalar fields.  We show that in these models the
infra-red (IR) fixed points are located on the boundary of ${\cal M}$,
where the space-time metric becomes singular.
\end{abstract}


\newpage


Over the past year  a lot of attention has been given to elucidate, 
via AdS/CFT correspondence, 
the renormalization group (RG) flow of strongly coupled gauge theory
in terms of static domain wall solutions in anti-deSitter (AdS)
supergravity theories \cite{270,040,020,620}.  In particular, the kink
solutions of scalar-fields interpolating between the (supersymmetric
AdS) extrema of gauged supergravity potentials provide a frame-work to
address the aspects of RG flows on the dual gauge theory.  (The first
examples of supersymmetric domain wall solutions was found in the
context of D=4 N=1 supergravity theory \cite{030} and reviewed in
\cite{210}.)

In gauged supergravity theories the scalar potential is of a restricted
type, in particular, most of the supersymmetric extrema are {\it maxima}
of the potential and thus in general it is difficult to obtain scalar
kink solutions which at short space-time distances asymptote to the
Cauchy horizon of the AdS space-time and which would correspond to
infra-red (IR) fixed points. For this latter type of behavior in the IR
the potential exhibits the second supersymmetric extremum, which is a
saddle point or a minimum. Nevertheless an example of this type has
been discussed in \cite{040} for non-Abelian gauged supergravity,
associated with the massless supermultiplets of Type IIB supergravity
compactified on a five-sphere $S^5$ (D=5 N=8 gauged supergravity). In
\cite{290} another example of this type is due to the scalars of a
massive supermultiplet of sphere reductions -- those are breathing modes
parameterizing the volume of the internal sphere.

In general, the maximally supersymmetric extremum of gauged supergravity
potentials, which is always a maximum, is responsible for the kink
solutions  that  asymptote at large distances (the ultra-violet (UV) fixed
point) to the boundary of the AdS space-time, i.e. the corresponding
conformal field theory in the UV is specified at the boundary of AdS
space-time. On the other hand, the flows from this UV fixed point
generically exhibit a singularity in the IR,  i.e. the kink solution
approaches at short distances the values of the run-away potential. For
such examples within D=5 N=8 gauged supergravity see, e.g.,
\cite{140,280}.

These latter features of the RG flows seem to be more generic within a
set-up of gauged supergravity theories. For N=2 D=5 gauged
supergravity with vector supermultiplets only, it has been shown
\cite{140,280} that the potential has only a single extremum for any
physical domain ${\cal M}$ of the moduli space for the scalar fields
($\phi_A$) and this extremum corresponds to an UV fixed point, where
the space-time asymptotes to the AdS boundary. The same holds
\cite{280} for N=2 D=4 gauged supergravity with vector
supermultiplets, only. On the other hand the kink solution at small
distances (in the IR regime) necessarily reaches the singular domain
at the boundary of $\cal M$, where both the scalar potential and the
space-time become singular. Flows into such singular regions have been
also discussed in \cite{310,320,610}.

The purpose of this letter is to show that, in spite of the singular
nature at the boundary of ${\cal M}$, the solutions always exhibit an
IR fixed point there, i.e. the $\beta$ functions vanish there
($\beta^A\Big|_0=0$) and their first derivatives $\partial_A
\beta^B(\phi)\Big|_0$ are positive definite.  The ordinary space-time
(in D=4,5) also becomes singular at this (small) distance.  However,
the geodesic distance on ${\cal M}$, is {\it infinite}.  To reiterate,
these results hold for D=5 and D=4 gauged supergravity with eight
supercharges and with vector supermultiplets, only. In D=4, the
results change \cite{210,030} if one allows for only four unbroken
supercharges.

Let us start with some general remarks. In order to get a proper RG
flow interpretation of the supergravity we have to choose a specific
coordinate system for the space-time metric \cite{350,020, 330}:
\be010
ds^2 = u^2 \Big( -dt^2 + d\vec x^2 \Big) + { du^2 \over W(u)^2 u^2},
\ee
where $W(u)^2$ becomes the cosmological constant or inverse AdS radius at
the UV fixed point ($u \rightarrow \infty$), however, for a finite $u$,
it varies.  The IR region corresponds to $u \rightarrow 0$. Using
this coordinate system, the solution preserves supersymmetry if the
scalar fields satisfy the following equation \cite{020,330}:\footnote{The
notation is for real scalars, in any dimension D. For complex scalars in
D=4 the equations require minor modifications. See later. The form of the
metric (\ref{010}) and the $\beta$ function equation (\ref{020}) are a
consequence of the first order differential equations -- the Killing spinor
equations.
For D=4, see \cite{210}, for D=5, see e.g., \cite{040}.}
\be020
\beta^A \equiv u {d \over du} \phi^A = g^{AB} \partial_B \log|W|^{-(D-2)}
= g^{AB} \partial_B \log C ,
\ee
where $\phi^A$ are the scalar fields and the second equation is a
proposal \cite{270,040,020} for the definition of the $C$-function.
The quantities $W$ and $g_{AB}$ , that enter the metric (\ref{010})
and the $\beta^A$ functions (\ref{020}), are nothing but the
superpotential and the metric of the scalar fields $\phi^A$ whose
Lagrangian reads
\be030
S_D = \int d^D x\ \Big[{\frac{1}{2}} R - V - {\frac{1}{2}} 
g_{AB} \partial_{\mu} \phi^A \partial^\mu \phi^B \Big] ,
\ee 
and the potential has the form
\be040
V = 2(D-2)(D-1) \, \Big( \, {D-2 \over D-1} g^{AB} 
\partial_A W \partial_B W -  W^2 \, \Big)\ .
\ee

Obviously, at extrema of $W$ the $\beta^A$ functions (\ref{020}) vanish
and we reach an AdS space. In general, the constraints of gauged
supergravity impose constraints on the form of $W$. Thus only very
specific solutions of the RG flows can take place.\footnote{In principle,
one may choose $W$ to be any function of $\phi^A$, and thus break
supersymmetry; nevertheless this case would still correspond to a stable
solution \cite{080,360}. However, if one insists on an embedding in a
supergravity with eight supercharges, $W$ is of a restricted form.}

Let us focus on D=5 case, first.
D=5 gauged supergravity with eight supercharges and the vector
super-multiplets, whose scalars $\phi^A$ parameterize a hypersurface
${\cal M}$,  are defined by a cubic equation \cite{120}:
\be050
F(X) \equiv {1 \over 6} \, C_{IJK} X^I X^J X^K = 1, 
\ee
where $I = 0,1,2, \ldots n$ and $n$ is the number of auxiliary vector
supermultiplets $X^I$ and $C_{IJK}$ are the coefficient defining the
cubic Chern-Simons term in the supergravity Lagrangian.  In Calabi-Yau
compactifications these are the topological intersection numbers and the
constraint (\ref{050}) means that the volume is kept fix. In general
${\cal M}$ is not connected and consists of different branches, separated
by regions where $F(X) < 0$; see figure 1.

\begin{figure}  \label{fig1}
\begin{center}
\includegraphics[angle = -90,width=60mm]{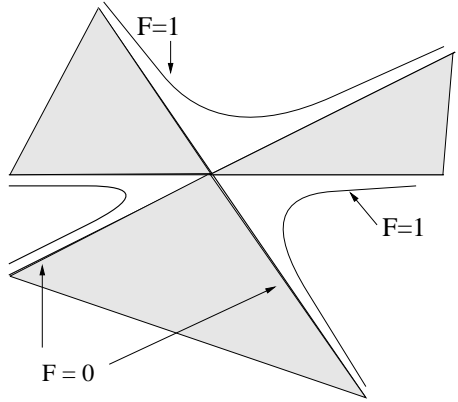}
\end{center}
\caption{The scalar fields of vector super-multiplets of D=5 theory
parameterize a manifold that consists of different branches. The
straight lines correspond to F=0 domain and shaded areas to $F < 0$
domains.}
\end{figure}

The potential is generated by gauging a $U(1)$ subgroup of the $SU(2)$
R-symmetry \cite{110} and the superpotential $W$ and the scalar metric
have the form
\be060
W = \alpha_I X^I \qquad , \qquad \ g_{AB}= - {1 \over 2}  \Big(\partial_A X^I 
\partial_B X^J\, \partial_I \partial_J \log F(X)\Big)\Big|_{F=1}  \ ,
\ee
where $\alpha_I$ are  constant parameters of the  $U(1)$ gauging. 
Using general formulae from \cite{120}, one can expand $W$ 
around a critical point ($\partial_A W =0$) and one finds
\be070
{W \over W_0} = 1 + {1 \over 3} g_{AB} 
(\phi^A- \phi^A_0)(\phi^B -\phi_0^B) + ...  
\ee
Inserting this expression into the $\beta$-function (\ref{020}) one gets
\cite{020}
\be080
\beta^A = -2 (\phi^A - \phi_0^A) + {\cal O}(\Delta\phi^2), 
\ee
and consequently $\partial_B \beta^A = -2 \, \delta_B^{\ A}$, which means
that extrema of $W$ are UV fixed points \cite{140}.  In fact the sign in
front of the second term in the expansion (\ref{070}) determines the
nature of the fixed point; for positive values it is an UV and for
negative an IR fixed point. Since there is only one extremum of $W$ per
nonsingular physical domain of $\cal M$ \cite{280} \footnote{The physical
domain is specified by the constraint that the scalar metric remains
positive-definite in the domain of $\cal M$. Same conclusions are
obtained by imposing the convexity constraint of the internal space
\cite{410}.}, one concludes that this extremum always
corresponds to the UV fixed point. Equivalently, space-time asymptotes to
the AdS boundary there \cite{140, 280}.

Having this UV fixed point, let us now discuss the domain near the
boundary of ${\cal M}$, which is defined by zeros of $F(X)$. For the
discussion,
it is convenient to use projective coordinates, i.e.\ to define the
physical scalars as $\phi^A = X^A/X^0$ and $F(X)$ and the
superpotential becomes
\be090
F = (X^0)^3 p_3(\phi),  \qquad 
W = X^0\Big( \alpha_0  + \alpha_A \phi^A \Big), 
\ee
where $p_3(\phi)$ is a polynomial of third degree in $\phi^A$. For the
single scalar case, e.g., we obtain $p_3(\phi) = (\phi - \phi_1)
(\phi- \phi_2)(\phi - \phi_3)$ and reach the boundary of ${\cal M}$ at
zeros of this polynomial, i.e.\ at $\phi = \phi_{1,2,3}$. Due to the
requirement of $F=1$ the zeros of $p_3(\phi)$ translate into poles of
$X^0$ and therefore into poles of $W$. There are two cases to
distinguish, if $p_3(\phi)$ has a single ($n=1$) or double ($n=2$)
zero. Note the case $\phi_1 = \phi_2 = \phi_3$ is trivial; there are
no physical scalars. The situation for the multi-scalar case is
analogous, i.e.\ we have either a linear or quadratic zero of $p_3$
(and the corresponding pole of $W$).  Therefore, for generic values of
$\alpha_I$ the superpotential behaves near the boundary as
\be100
W \sim \lambda^{-n/3}, \ \  n=1,2,
\ee
where $\lambda \rightarrow 0$ denotes a scaling parameter in
terms of the deviation of the scalars from their boundary value, e.g., \
for the single scalar case we can write $\lambda\equiv \phi - \phi_0$.
Similarly, one obtains for the scaling of  the metric
\be110
g_{AB} = {1 \over 3 \lambda^2} \hat g_{AB} \ ,
\ee
where $\hat g_{AB}$ is the part which remains finite (and positive
definite) on the boundary. Using these simple scaling arguments one
obtains for the $\beta$ function near the boundary
\be120
\beta^A = 3n \hat g^{AB} (\phi^B - \phi_0^B), \ \   n=1,2.
\ee
In the case where all $\beta$ functions vanish, we reach a fixed point of
the RG flow and because $\partial_B \beta^A$ is positive definite, these
are indeed IR fixed points. This result should be  contrasted with examples
where there is  a regular UV fixed points in the bulk 
(with a  non-singular AdS space-time); see (\ref{080}). 

Moreover, the leading order term of the $\beta$ function
near the UV fixed point was universal. On the other hand the $\beta$
function near the boundary depends on $\hat g^{AB}$, and is thus
model-dependent. In the following we consider a number of examples.

\noindent{\it Single scalar case.} Inserting the near-boundary value
of $\lambda=\phi-\phi_0$, we obtain $\beta = 3\, n\, c \, (\phi -
\phi_0)$ with some model-dependent positive $c$. As a
solution for the scalar field $\phi$ and $W$ one obtains \cite{010}
\be132
\phi -\phi_0 \sim u^{3\, n\, c} \quad, \quad W \sim u^{-n^2\,c}.
\ee
For $c=1$ the metric can be written as \footnote{The complete
analytic metric can be obtained by using the results of \cite{020}.}
\be134
\ba{ll}
ds^2 = (a z - z_0)^2 \Big( -dt^2 + d\vec x^2 \Big) + 
dz^2, \qquad & {\rm for:}
\quad n =1 ,\\
ds^2 = \sqrt{a z - z_0} \Big( -dt^2 + d\vec x^2 \Big) + dz^2, & {\rm for:}
\quad n =2, 
\ea
\ee
where $a$ is basically an integration constant. In both cases the
space-time metric is singular. The second case has been encountered also
in \cite{310,320}. Since this singularity corresponds to the boundary of
${\cal M}$, it is an infinite geodesic distance away, i.e.\ $\int_{\phi
=\phi_0} \sqrt{g_{\phi\phi}}\, d\phi = \infty$ (see also the example
below), even though the affine parameter along this trajectory as
measured by the space-time radius $z$ or $u$ remains finite.
Moreover, the zeros of $F(X)$ are not really the ``end of the world'',
they rather separate different phases of the theory related to the
different branches of ${\cal M}$ and the infinite geodesic distance
may indicates that each branch is related to a different topology.
{From} the field theory point of view the different branches are related
to different sides of an IR fixed point.

In order to elucidate the implications of this singularity from the
higher dimensional viewpoint one may try to adopt arguments as used in
\cite{340} to regulate the solution or techniques used to resolve
conical singularities in Calabi-Yau compactifications~\footnote{We
thank J.\ Schwarz for a comment on this.}. E.g., if we replace the
flat world-volume in the case $n=1$ by an space of constant curvature
$k$, i.e.\ a de Sitter space, the D=5 metric becomes flat iff $k
=a^2$. On the other hand, examples with $n=2$ that arise in D=4,5
theories as sphere reductions of M/string-theory, e.g., STU model
(discussed later) have a higher dimensional interpretation as
distributions of positive tension branes (see e.g., \cite{291,610} and
references therein), and thus from higher dimensional perspective do
not seem to suffer from pathologies.  Let us also mention that higher
curvature corrections provide a natural cut-off for the volume of the
internal space and may be used as a regulator, see
\cite{200}. However, one has to keep in mind that any curvature
cut-off will also remove the IR fixed point! The $\beta$ functions
vanish only {\em on} the singularity.

\bigskip

Let us further consider  two other specific examples:
\newline
{\it (i)} $F = ST^2 -b \, T^3$, which is a single scalar case, 
\newline
{\it (ii)} $F= STU$, which has two scalars.

{\it Case (i).} Defining the physical scalar as $\phi = {T \over S}$,
which parameterizes the angle in figure 1, we reach the boundaries at
$F= S^3 \phi^2 (1 - b\, \phi)=0$, i.e.\ at $\phi \rightarrow 0,
{1/b}$. The first case corresponds to a double zero ($n=2$) and the
latter to a single zero ($n=1$).  The inverse metric and the
derivative of the superpotential $W = S + T= S(1 + \phi)$ (taking
$\alpha_I = (1,1)$) become
\be130
g_{\phi\phi} = {1 \over 3\, \phi^2 (1-b \, \phi)^2 } 
\quad , \quad
{\partial_{\phi} W \over W} =  {(3b+1) \phi - 2 \over
3 \phi (\phi+1) (1- b \, \phi)} ,
\ee
and therefore the $\beta$ function reads
\be140
\beta^{\phi} = - 3 \, {\phi (1- b\, \phi)\, \big[(3\, b +1) \, \phi 
-2 \big] \over  \phi +1}.
\ee
We find three fixed points 
\be150
\phi_{UV} = {2 \over 3 b +1} 
\quad , \quad \phi_{IR} = \{0\, , \, 1/b\}, 
\ee
with $\partial_{\phi} \beta|_{UV} = -2$ and $\partial_{\phi} \beta|_{IR}
= \{6, 3/b^2\}$ for $\phi = \{0,1/b\}$, respectively; notice the
different values for the two IR fixed points which  were in general  
parameterized by the coefficient $c$ in (\ref{132}). The $\beta$
function is shown in figure 2 and it behaves smoothly over all branches
of ${\cal M}$.

\begin{figure}  \label{fig2}
\begin{center}
\includegraphics[angle = -90,width=70mm]{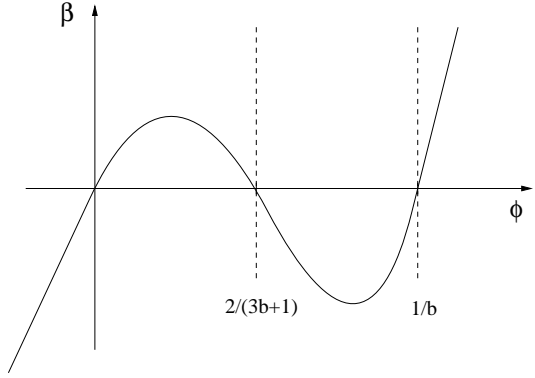}
\end{center}
\caption{This figure shows the $\beta$ function (\ref{140}) for
a single scalar with which has three fixed points: $\phi_{IR} =
0,1/b$ and $\phi_{UV} = 2/(3b+1)$. The IR fixed points lie on the 
boundary of ${\cal M}$.}
\end{figure}

{\it Case (ii).} In this case we take as physical scalars $(T,U)$
($S=1/TU$) and for the superpotential we take again $\alpha_I =(1,1,1)$
or $W = 1/TU + T +U$. We find for the $\beta$ functions
\be160
\beta^T = {2 T \Big(1 - 2 T^2 U + T U^2\Big) \over
           1 + T^2 U + T U^2} \quad , \quad
\beta^U = {2 U \Big(1 - 2 T U^2 + T^2 U\Big) \over
           1 + T^2 U + T U^2} \ .
\ee
There are only two fixed points:
\be170
\Big(T , U\Big)_{UV} = (1,1) \qquad , \qquad
\Big(T , U\Big)_{IR} = (0,0), 
\ee
with $\partial_A \beta^B\Big|_{UV} = -2 \, \delta_A^{\ B}$ and $\partial_A
\beta^B\Big|_{IR} = 2 \, \delta_A^{\ B}$. Therefore,  only 
one point on the boundary of $\cal M$ is an IR fixed point.  However, this
is a very special example of the two-scalar case. If one modifies $F$,
e.g., \ $F = STU \rightarrow STU - b \, U^3$,
one can obtain more than one  IR fixed point.

\bigskip

We now comment also on D=4  gauged supergravity with eight supercharges.
In this case, we have to replace the coordinates $X^I$ by the
symplectic section $(X^I , F_I)$ where $F_I = \partial_I F(X)$
denotes the derivative of the prepotential. Using this symplectic
section, a real superpotential can be defined by
\be180
\widehat W \equiv {1 \over 2} \, \xi\, |W e^{K/2}| = 
{1 \over 2} \, \xi \, | \alpha_I X^I - \beta^I F_I| e^{K/2}\ ,
\ee
with $W$ and $K$ as Super- and K\"ahler potential, respectively and
$\xi = \pm 1$ (it can change sign iff $W$ crosses zero).  In our
notation the K\"ahler metric is defined as $g_{A\bar B} = {1 \over 2}
\partial_A \partial_{\bar B} K$. Again, extrema of $\widehat W$
correspond to UV fixed points \cite{280}. (For notation and
conventions, see \cite{630}.) Here let us turn to the investigation of
the boundary behavior of the RG equations.

In D=5 case one had to look at poles in $X^0$, see eq.\
(\ref{090}). In D=4 these singularities are equivalent to poles in
$e^{K/2}$.  Taking a generic flux vector $(\alpha_I , \beta^I)$ the
$\beta$ functions near such poles take the form:
\be190
\beta^A = - g^{A \bar B} (\partial_{\bar B} K + {\rm finite\ term} ) \ .
\ee
As before, we can use general scaling arguments to verify the
behavior: $e^{-K} \sim \lambda^{n}$ and $g_{A \bar B} \sim {n \over 2}
\lambda^{-2}$, where $n$ is again the degree of the pole with $n=1,2$.
 
An equivalent relation to (\ref{120}) for the behavior 
near the pole $(z^A - z^A_0)
\simeq 0$  takes the form:
\be200
\beta^A = 2\, \bar g^{A \bar B} (z^B - z_0^B) \ , 
\ee
where  $z^B = {X^B \over X^0}$ is the complex scalar. Because the derivatives
of these $\beta$ functions are  positive definite we  again 
obtain  the IR fixed points
at $z^A \simeq z_0^A$. 

As an example, let us consider again  the
single-scalar case given by the prepotential $F = X^0 X^1$, which yields the 
K\"ahler potential $e^{-K} = -i(T - \bar T)$. In this case the $\beta$
function near the zero of $e^{-K}$ becomes
\be210
\beta(T) = 2 (T -\bar T)\ , 
\ee
and therefore  one reaches an IR fixed point at: Re$T=const.$, Im$T =0$.
As for the UV fixed point, with the choice   $(\alpha_I , \beta^I) =
(1,1,0,0)$, one find the   extremum of $\widehat W = | 1-i\, T|e^{K/2}$ at 
Re$T=0$, Im$T=1$;  near 
this point the  $\beta$ function  is of the form: 
\be777
\beta(T) = -2 (T-i)\ ,
\ee
with  its first derivative  negative and thus identified as a UV fixed point.



\vspace{1cm}

{\bf Acknowledgments} 

We would to thank H. L\" u, D.\ Mini\'c and  H.\ Verlinde for useful
discussions.  M.C. would like to thank Caltech Theory Group  for hospitality.
The work is supported by a DFG Heisenberg grant (K.B.), in
part by the Department of Energy under grant number DE-FG03-92-ER 40701
(K.B.), DOE-FG02-95ER40893 (M.C.) and the University of Pennsylvania
Research Foundation (M.C).


%
%

\providecommand{\href}[2]{#2}\begingroup\raggedright\endgroup


\end{document}